\documentstyle[12pt]{article}

\global\arraycolsep=2pt
\makeatletter
\def\fmslash{\@ifnextchar[{\fmsl@sh}{\fmsl@sh[0mu]}}
\def\fmsl@sh[#1]#2{%
  \mathchoice
    {\@fmsl@sh\displaystyle{#1}{#2}}%
    {\@fmsl@sh\textstyle{#1}{#2}}%
    {\@fmsl@sh\scriptstyle{#1}{#2}}%
    {\@fmsl@sh\scriptscriptstyle{#1}{#2}}}
\def\@fmsl@sh#1#2#3{\m@th\ooalign{$\hfil#1\mkern#2/\hfil$\crcr$#1#3$}}
\makeatother

\begin{document}

\vspace{1.cm}
\begin{center}
\Large\bf  $\rho$-Decay Widths of Excited Heavy Mesons From\\ 
Light-Cone QCD Sum Rules\\
 in the Leading Order of HQET 
\end{center}
\vspace{0.5cm}
\begin{center}
{Shi-Lin Zhu and Yuan-Ben Dai}\\\vspace{3mm}
{\it Institute of Theoretical Physics,
 Academia Sinica, P.O.Box 2735, Beijing 100080, China }
\end{center}

\vspace{1.5cm}
\begin{abstract}
The couplings and decay widths of the processes  
$(1^+,2^+) \to (0^-, 1^-) +\rho $ are studied with the light-cone 
QCD sum rules in the leading order of heavy quark effective
theory. These processes are employed to estimate the two pion
transition widths of the $(1^+,2^+)$ doublet through the 
low mass tail of the $\rho$ resonance.
The ambiguity due to the presence of two distinct $1^+$ states is solved.
Our calculation shows that 
the two pion decay widths of the $(1^+, 2^+)$ doublets are 
much smaller than the single pion decay widths. 
However, the $B_1$, $B_2^\ast$ mesons should 
also have significant two pion decay widths around $1.5$MeV.
\end{abstract}


\pagenumbering{arabic}

\section{ Introduction}
\label{sec1} 
The heavy quark effective theory (HQET) \cite{grinstein} provides a 
systematic expansion of the heavy hadron spectra and transition amplitude
in terms of $1/m_Q$, where $m_Q$ is the heavy quark mass.
The spectrum pf the ground state heavy meson has been studied with 
the QCD sum rules \cite{svz} in HQET in \cite{col91}.
In \cite{huang} the mass of the lowest excited heavy meson 
doublets $(1^+,2^+)$ and $(0^+,1^+)$ were studied with QCD sum rules in the
heavy quark effective theory (HQET) up to the order of ${\cal O}(1/m_Q)$. 

The QCD sum rules was used to analyse the exclusive radiative 
$B$-decays with the help of the light-cone vector meson wave function 
in \cite{braun94}. With the same formalism the off-shell $g_{B^* B\rho}$ 
and $g_{D^* D\rho}$ couplings were calculated in \cite{aliev96}.
In this work we employ the light-cone QCD sum rules 
(LCQSR) in HQET to calculate the 
decay widths of the processes $(1^+,2^+) \to (0^-, 1^-) +\rho$
to the leading order of $1/ m_Q$.
When the LCQSR is used to calculate the coupling constant, the 
double Borel transformation is always invoked so that the excited states and 
the continuum contribution can be treated quite nicely. 

One difficult problem encountered in studying the decay 
widths of excited heavy mesons is due to the degeneracy of the spectra.
There are a pair of states for any spin-parity $j^P$ with
close values in their masses but quite different in magnitudes of their
decay widths, except for the ground states.
Only in the $m_Q\to\infty$ limit, is there a conserved quantum 
number $j_{\ell}$, the angular momentum of the light component, 
which can be used to differentiate the two states. Therefore, 
HQET has important and unique advantage for this purpose. 
These are the motivation for our approach of using LCQSR in HQET.

The proper interpolating current $J_{j,P,j_{\ell}}^{\alpha_1\cdots\alpha_j}$
for the states with the quantum number $j$, $P$, $j_{\ell}$ in HQET was
given in \cite{huang}. They were proved to satisfy the following conditions 
\begin{eqnarray}
\label{decay}
\langle 0|J_{j,P,j_{\ell}}^{\alpha_1\cdots\alpha_j}(0)|j',P',j_{\ell}^{'}\rangle&=&
f_{Pj_l}\delta_{jj'}
\delta_{PP'}\delta_{j_{\ell}j_{\ell}^{'}}\eta^{\alpha_1\cdots\alpha_j}\;,\\
\label{corr}
i\:\langle 0|T\left (J_{j,P,j_{\ell}}^{\alpha_1\cdots\alpha_j}(x)J_{j',P',j_{\ell}'}^{\dag
\beta_1\cdots\beta_{j'}}(0)\right )|0\rangle&=&\delta_{jj'}\delta_{PP'}\delta_{j_{\ell}j_{\ell}'}
(-1)^j\:{\cal S}\:g_t^{\alpha_1\beta_1}\cdots g_t^{\alpha_j\beta_j}\nonumber\\[2mm]&&\times\:
\int \,dt\delta(x-vt)\:\Pi_{P,j_{\ell}}(x)
\end{eqnarray}
in the $m_Q\to\infty$ limit, where $\eta^{\alpha_1\cdots\alpha_j}$ is the
polarization tensor for the spin $j$ state, $v$ is the velocity of the heavy
quark, $g_t^{\alpha\beta}=g^{\alpha\beta}-v^{\alpha}v^{\beta}$ is the
transverse metric tensor, ${\cal S}$ denotes symmetrizing the indices and
subtracting the trace terms separately in the sets $(\alpha_1\cdots\alpha_j)$
and $(\beta_1\cdots\beta_{j})$, $f_{P,j_{\ell}}$ and $\Pi_{P,j_{\ell}}$ are
a constant and a function of $x$ respectively which depend only on $P$ and $%
j_{\ell}$. Because of equations (\ref{decay}) and (\ref{corr}), the sum rule
in HQET for decay widths derived from a correlator containing such currents
receive no contribution from the unwanted states with the same spin-parity
as the states under consideration in the $m_Q\to\infty$. 

%%%%%%%%%%%%%%%%%%%%%%%%%%%%%%%%%%%%%%%%%%%%%%%%%%%%%%%%%%%%%%%%%%%%%%%%%%%%%%%%
\section{ Sum rules for decay amplitudes}

In the present work we shall confine ourselves to 
the leading order of $1/m_Q$ expansion. Denote the doublet 
$(1^+,2^+)$ with $j_{\ell}=3/2$ by $(B_1,B_2^*)$. 
From covariance and conservation of the angular momentum of the light 
component in the $m_Q\to\infty$ limit, there are three independent decay 
amplitudes for the decays of $B_1$, $B_2^*$ to the
rho meson and the ground states $B$, $B^*$ in the doublet $(0^-,1^-)$ 
with $j_{\ell}=1/2$. They are characterized by $(j_h, l)=(2,2)$, 
$(1,2)$, and $(1,0)$, where $l$ and $j_h$ are the 
orbital and total angular momentum of the rho meson 
respectively \cite{eichten93}. 
The S-wave amplitude is dominant.

The amplitudes read as follows:
\begin{equation}
\label{coup1}
\begin{array}{ll}
 M(B_1\to B^*\rho)=&I\; e^\ast_\mu\{ 
 i\epsilon^{\mu\nu\beta\sigma} v_\sigma  
( \eta^\alpha \epsilon^\ast_\beta +\eta_\beta {\epsilon^\ast}^\alpha ) 
(q^t_\nu q^t_\alpha -{1\over 3} g_{\nu\alpha}^t q_t^2) g_{(2,2)}(B_1, B^\ast )\\
&+ i\epsilon^{\mu\nu\beta\sigma} v_\sigma  
( \eta^\alpha \epsilon^\ast_\beta -\eta_\beta {\epsilon^\ast}^\alpha ) 
(q^t_\nu q^t_\alpha -{1\over 3} g_{\nu\alpha}^t q_t^2) g_{(1,2)}(B_1, B^\ast )\\
&+i\epsilon^{\mu\alpha\beta\sigma}\eta_\alpha\epsilon^\ast_\beta v_\sigma  
 g_{(1,0)} (B_1, B^\ast) \} \; ,
\end{array}
\end{equation}
\begin{equation}
\label{coup2}
M(B_1\to B\rho)=
I\; \eta_\beta  e^\ast_\alpha 
[(q_t^\alpha q_t^\beta-{1\over 3}q^2_t g_t^{\alpha\beta}) g_{(2,2)}(B_1, B) 
+g_t^{\alpha\beta}g_{(1,0)}(B_1,B) ]\;,
\end{equation}
\begin{eqnarray} 
\label{coup3} \nonumber
 M(B^*_2\to B^*\rho)=&I\; \eta^{\alpha_1}_{\alpha_2}{\epsilon^\ast}^\beta 
 (g^t_{\beta\mu}  e^\ast_{\alpha_1} -e^\ast_\beta g^t_{\alpha_1\mu})
 (q^\mu_tq^{\alpha_2}_t-{1\over 3} g_t^{\mu\alpha_2}q_t^2)g_{(1,2)}(B_2^*, B^*) \\
 &+\eta_{\alpha\beta}{\epsilon^\ast}^\alpha {e^\ast}^\beta g_{(1,0)}(B_2^*, B^*)
\; ,
\end{eqnarray}
\begin{equation} 
 \label{coup4}
M(B^*_2\to B\rho)=I\; \eta_{\alpha_1\alpha_2} 
i\epsilon^{\mu\nu\sigma\alpha_1} e^\ast_\mu q^t_\nu v_\sigma 
 q^t_{\alpha_2} g_{(1,2)}(B^*_2, B)\;,
\end{equation} 
where $\eta_{\mu\nu}$, $\eta_{\mu}$, $\epsilon_{\mu}$ and 
$e^{(\lambda )}_\mu$ are polarization tensors for 
the $2^+$, $1^+$, $1^-$ heavy mesons and the rho meson respectively. 
$q_{t\mu}=q_{\mu}-v \cdot qv_{\mu}$. $I=1$, ${1\over \sqrt{2}}$ 
for the charged and neutral rho meson respectively. 
The amplitude $g_{(2,2)}$, $g_{(1,2)}$ and $g_{(1,0)}$ in 
(\ref{coup1})-(\ref{coup4})
corresponds to the $(j_h, l)=(2,2)$, $(1,2)$, and $(1,0)$ decay amplitude respectively.

Due to the heavy quark symmetry the D-wave and S-wave amplitudes 
in these four processes are related. From vector current conservation 
we have another constraint. For example, 
vector current conservation requires that the tensor structure be 
$(q_t^\alpha q_t^\beta-q^2_t g_t^{\alpha\beta})$
for the process $B_1\to B^\ast\rho$. From this condition we have
$g_{(1,0)}(B_1, B)=-{2\over 3}q_t^2g_{(2,2)}(B_1, B)$.
In HQET there exist only two independent coupling constants 
$g_d=g_{(1,2)}(B^*_2, B)$ and $g_s=g_{(1,0)}(B_2^*, B^*)$, 
corresponding to the D-wave and S-wave decays respectively. 

In order to derive the sum rules for the coupling constants $g_d$ and $g_s$
we consider the correlators 
\begin{eqnarray}
\label{7a} \nonumber
 &\int d^4x\;e^{-ik\cdot x}\langle\rho (q)|T\{J^{\beta}_{1,-,\frac{1}{2}}(0)
 J^{\dagger\alpha}_{1,+,\frac{3}{2}}
 (x)\}|0\rangle =
I\; e^\ast_\mu\{ 
[ i\epsilon^{\mu\nu\beta\sigma} v_\sigma  
 (q^t_\nu q^t_\alpha -{1\over 3} g_{\nu\alpha}^t q_t^2)\\ \nonumber
& +(\alpha \leftrightarrow \beta )] G_{(2,2)}^{B_1 B^\ast}(\omega,\omega') 
+[ i\epsilon^{\mu\nu\beta\sigma} v_\sigma  
(q^t_\nu q^t_\alpha -{1\over 3} g_{\nu\alpha}^t q_t^2) 
 -(\alpha \leftrightarrow \beta )] G_{(1,2)}^{B_1 B^\ast}(\omega,\omega') 
\\ &
+i\epsilon^{\mu\alpha\beta\sigma} v_\sigma 
G_{(1,0)}^{B_1 B^\ast}(\omega,\omega') \}
\;,
\end{eqnarray}
\begin{eqnarray}
\label{7b}\nonumber
&\int d^4x\;e^{-ik\cdot x}\langle\rho (q)|T\{J_{0,-,\frac{1}{2}}(0)
 J^{\dagger\beta}_{1,+,\frac{3}{2}}(x)\}|0\rangle = 
I\;{e^\ast}^\beta [(q_t^\alpha q_t^\beta-{1\over 3}q^2_t g_t^{\alpha\beta})
G_{(2,2)}^{B_1  B}(\omega,\omega') \\
&+g_t^{\alpha\beta}G_{(1,0)}^{B_1  B}(\omega,\omega') ]
\;,
\end{eqnarray}
\begin{eqnarray}
\label{7c}\nonumber
& \int d^4x\;e^{-ik\cdot x}\langle\rho (q)|T\{J^{\beta}_{1,-,\frac{1}{2}}(0)
 J^{\dagger\alpha_1\alpha_2}_{2,+,\frac{3}{2}}(x)\}|0\rangle = 
I\;\{ (g^t_{\beta\mu}  e^\ast_{\alpha_1} -e^\ast_\beta g^t_{\alpha_1\mu})
 (q^\mu_tq^{\alpha_2}_t \\
&-{1\over 3} g_t^{\mu\alpha_2}q_t^2)
 G_{(1,2)}^{B_2^*  B^*}(\omega,\omega')
+{e^\ast}^{\alpha_2} g_t^{\beta\alpha_1}
G_{(1,0)}^{B_2^*  B^*}(\omega,\omega') \}\;,
\end{eqnarray}
\begin{eqnarray}
\label{7d}
 \int d^4x\;e^{-ik\cdot x}\langle\rho (q)|T\{J_{0,-,\frac{1}{2}}(0)
 J^{\dagger\alpha_1\alpha_2}_{2,+,\frac{3}{2}}(x)\}|0\rangle = & 
I\; i\epsilon^{\mu\nu\sigma\alpha_1} e^\ast_\mu q^t_\nu v_\sigma 
 q^t_{\alpha_2} G_{(1,2)}^{B^*_2 B}(\omega,\omega')\;,
\end{eqnarray}
where $k^{\prime}=k-q$, $\omega=2v\cdot k$, $\omega^{\prime}=2v\cdot
k^{\prime}$, $q_t^2=q^2-(q\cdot v)^2$ and $q^2=m_\rho^2$. 

The interpolationg currents are given in \cite{huang} as 
\begin{eqnarray}
\label{curr1}
&&J^{\dag\alpha}_{1,+,{3\over 2}}=\sqrt{\frac{3}{4}}\:\bar h_v\gamma^5(-i)\{
{\cal D}_t^{\alpha}-\frac{1}{3}\gamma_t^{\alpha}\fmslash{\cal D}_t\}q\;,\\
\label{curr2}
&&J^{\dag\alpha_1,\alpha_2}_{2,+,{3\over 2}}=\sqrt{\frac{1}{2}}\:\bar h_v
\frac{(-i)}{2}\{\gamma_t^{\alpha_1}{\cal D}_t^{\alpha_2}+
\gamma_t^{\alpha_2}{\cal D}_t^{\alpha_1}-{2\over 3}g_t^{\alpha_1\alpha_2}
\fmslash{\cal D}_t\}q\;,\\
\label{curr3}
&&J^{\dag\alpha}_{1,-,{1\over 2}}=\sqrt{\frac{1}{2}}\:\bar h_v\gamma_t^{\alpha}
q\;,\hspace{1.5cm} J^{\dag\alpha}_{0,-,{1\over 2}}=\sqrt{\frac{1}{2}}\:\bar h_v\gamma_5q\;,
\end{eqnarray}
where $h_v$ is the heavy quark field in HQET and $\gamma_{t\mu}=\gamma_%
\mu-v_\mu\fmslash v$.

Let us first consider the function $G_d(\omega,\omega')
\equiv G_{(1,2)}^{B^*_2 B}(\omega,\omega')$ in 
(\ref{7d}). As a function of two variables $\omega$ and $\omega^\prime$,
it has the following pole terms from double dispersion relation 
\begin{eqnarray}
\label{pole}
G_d(\omega,\omega')={f_{-,{1\over 2}}f_{+,{3\over 2}}g_d\over (2\bar\Lambda_{-,{1\over 2}}
-\omega')(2\bar\Lambda_{+,{3\over 2}}-\omega)}+{c\over 2\bar\Lambda_{-,{1\over 2}}
-\omega'}+{c'\over 2\bar\Lambda_{+,{3\over 2}}-\omega}\;,
\end{eqnarray}
where $f_{P,j_\ell}$ are constants defined in (\ref{decay}), 
$\bar\Lambda_{P,j_\ell}=m_{P,j_\ell}-m_Q$.

For deriving QCD sum rules we calculate the correlator (\ref{7d}) by the
operator expansion on the light-cone in HQET to the leading order of 
$1/ m_Q$. The expression with the tensor structure reads:
\begin{equation}\label{lam5}			
- \int_0^{\infty} dt \int  dx  e^{-ikx} 
\delta (-x-vt){\bf Tr} \{ (i\gamma_5 ){1+{\hat v}\over 2} (-i\gamma_5)
(D^\alpha_t -{1\over 3} \gamma^\alpha_t {\hat D}_t)
\langle \rho (q)| u(x) {\bar d}(0) |0\rangle   \} \; ,
\end{equation}

The rho wave function is 
defined as the matrix elements of nonlocal operators between the vacuum and 
rho meson state. Up to twist four the Dirac components of the rho meson 
wave function are \cite{braun94}:
\begin{equation}
\label{wf1}
<0| {\bar \psi} (0) \sigma_{\mu\nu} \psi (x) |\rho (p,\lambda )>
=i (e^{(\lambda )}_{\mu} p_\nu - e^{(\lambda )}_{\nu} p_{\mu} ) f^T_{\rho} 
\int_0^1 du \; e^{-iuqx} \varphi^T_{\rho}(u, \mu ) 
\end{equation}
\begin{equation}\label{wf2}
\begin{array}{ll}
<0| {\bar \psi} (0) \gamma_{\mu} \psi (x) |\rho (p,\lambda )>
=&i p_\mu { (e^{(\lambda )}x) \over (px)} f_{\rho} m_{\rho}
\int_0^1 du \; e^{-iuqx} (\varphi^L_{\rho}(u, \mu )  \\
& +\{ e^{(\lambda )}_\mu -p_\mu { (e^{(\lambda )}x) \over (px)} \}
f_\rho m_\rho \int_0^1 du \; e^{-iuqx} g^v_{\rho}(u, \mu )  
\end{array}
\end{equation}
\begin{equation}\label{wf3}
<0| {\bar \psi} (0) \gamma_{\mu}\gamma_5 \psi (x) |\rho (p,\lambda )>
=-{1\over 4} \epsilon_{\mu\nu\rho\sigma}e^{(\lambda ) \nu}p^\rho x^\sigma
 f_{\rho} m_{\rho} \int_0^1 du \; e^{-iuqx} g^a_{\rho}(u, \mu )  
\end{equation}
Due to the choice of the gauge  $x^\mu A_\mu(x) =0$, the path-ordered 
gauge factor \\
$P \exp\big(i g_s \int_0^1 du x^\mu A_\mu(u x) \big)$ has been omitted.
The wave functions $\varphi^T_{\rho} (u,\mu )$ and $\varphi^L_{\rho} (u, \mu )$
are the leading twist distributions in the fraction of total momentum 
carried by the quark in the transversely and longitudinally polarised 
rho meson respectively. Both wave functions $g^a_\rho (u,\mu )$ and 
$g^v_\rho (u,\mu )$ correspond to transverse spin distribution. 
They contain both twist-two and -three contributions. 
The twist-two contributions to the wave functions $g^v_\rho (u,\mu )$ and 
$g^a_\rho (u,\mu )$ can be expressed
in terms of the leading twist longitudinal wave function $\varphi^L_\rho (u, \mu )$:
\begin{equation}\label{gv1}
g_\rho^{v,\mbox{twist-2}} (u)={1\over 2} \{
\int_0^u dv {\varphi^L_\rho (u) \over 1-v }
+\int_u^1 dv {\varphi^L_\rho (u) \over v }\} \; ,
\end{equation}
\begin{equation}\label{gv2}
{d\over du}g_\rho^{a,\mbox{twist-2}} (u)=2 \{
-\int_0^u dv {\varphi^L_\rho (u) \over 1-v }
+\int_u^1 dv {\varphi^L_\rho (u) \over v }\}  \; .
\end{equation}
All the four wave functions $f=\varphi^T, \varphi^L, g^v, g^a$ are normalised
such that $\int_0^1 f(u) du =1$. 
We shall keep only the leading twist contributions, i.e., we use the 
asymptotic forms for the following wave functions:
\begin{equation}\label{l}
\varphi^L (u,\mu )=6u(1-u) \; ,
\end{equation}
\begin{equation}\label{gv3}
g^v (u,\mu )={3\over 4} (1+\xi^2) \; ,
\end{equation}
\begin{equation}\label{ga}
g^a (u,\mu )=6u(1-u) \; ,
\end{equation}
with $\xi =2u-1$. 
We use the wave function in \cite{dz} for $\varphi^T (u)$ with the mock hadron 
mass replaced by the average total quark energy \cite{ht}. 
\begin{equation}\label{t}
\varphi^T (u) =N exp\left( -{m^2\over 8\beta^2 u (1-u)} \right)
[\mu^2+\mu {\tilde \mu}+{\tilde\mu}^2 u(1-u)]
\;,
\end{equation}
where $N=0.338$, $\mu={m\over \beta}$, ${\tilde\mu} ={{\tilde m}\over \beta}$ 
with $m=330$MeV, ${\tilde m}=1130$MeV, and $\beta=320$MeV. 
The wave function (\ref{t}) satisfies the constraint from the QCD sum rule
analysis of the moments of the rho meson 
wave function \cite{chernyak84,braun94,sum}. The rho meson wave 
function is close to the asymptotic form, which was pointed out in 
\cite{chernyak84,dz,braun94}.
The rho meson coupling constants were determined from QCD sum rule analysis: 
$f_\rho^T =f_\rho =200$MeV \cite{chernyak84}. 

Expressing (\ref{lam5}) with the rho light-cone wave functions, we arrive at:
\begin{equation}\label{7dd}
G_d(\omega,\omega')= -{1\over 4} 
\int_0^{\infty} dt \int_0^1 du e^{i (1-u) {\omega t \over 2}}
e^{i u {\omega' t \over 2}}
u \{ f^T_\rho \varphi^T_{\rho}(u) -{it\over 4} f_\rho m_\rho g_\rho^a (u)
\} \; .
\end{equation}

Smilarly for the other D-wave functions in (\ref{7a})-(\ref{7c}),
we have:
\begin{equation}\label{r1}
G_{(2,2)}^{B_1 B^\ast}(\omega,\omega') =
{\sqrt{6}\over 4}G_d(\omega,\omega')\; ,
\end{equation}
\begin{equation}\label{r2}
G_{(1,2)}^{B_1 B^\ast}(\omega,\omega') =
{\sqrt{6}\over 12}G_d(\omega,\omega')\; ,
\end{equation}
\begin{equation}\label{r3}
G_{(2,2)}^{B_1 B}(\omega,\omega') =
{\sqrt{6}\over 6}G_d(\omega,\omega')\; ,
\end{equation}
\begin{equation}\label{r4}
G_{(1,2)}^{B_2^\ast B^\ast}(\omega,\omega') =
G_d(\omega,\omega')\; .
\end{equation}
For the S-wave functions, we have:
\begin{equation}\label{r5}
G_{(1,0)}^{B_1 B^\ast}(\omega,\omega') =
{\sqrt{6}\over 6}G_s(\omega,\omega')\; ,
\end{equation}
\begin{equation}\label{r6}
G_{(1,0)}^{B_1 B}(\omega,\omega') =
-{\sqrt{6}\over 3}G_s(\omega,\omega')\; ,
\end{equation}
where we define $G_s(\omega,\omega')\equiv
G_{(1,0)}^{B_1 B^\ast}(\omega,\omega')$.
\begin{equation}\label{r7}
G_s(\omega,\omega')={q_t^2\over 3}G_d(\omega,\omega')
={1\over 3} [m_\rho^2-{(\omega -\omega')^2\over 4}]G_d(\omega,\omega')
\; .
\end{equation}
From the above relations we know there are two independent coupling
constants only. This is in agreement with the discussions above.

In order to extract $g_d$ we first make Wick rotation, then make 
double Borel transformation 
to eliminate the single-pole terms in (\ref{pole}) . 
Subtracting the continuum contribution which is modeled by the 
dispersion integral in region 
$\omega ,\omega' \ge \omega_c$, we arrive at:
\begin{equation}\label{final-a}
 g_d f_{-,{1\over 2} } f_{+, {3\over 2} } = {1\over 2}
 e^{ { {\bar\Lambda}_{-,{1\over 2} } +{\bar\Lambda}_{+,{3\over 2} } \over T }}
 u_0\{ f^T_\rho \varphi^T_{\rho} (u_0)T(1-e^{-{\omega_c\over T}}) 
-{1\over 2} f_\rho m_\rho g_\rho^a(u_0) \}\;,
\end{equation}
where $u_0 ={T_1\over T_1+T_2}$, $T={T_1 T_2\over T_1+T_2}$, $T_1, T_2$
are the Borel parameters corresponding to the variables 
$\omega, \omega'$ respectively.
In obtaining (\ref{final-a}) we have used the Borel transformation formula:
${\hat {\cal B}}^T_{\omega} e^{\alpha \omega}=\delta (\alpha -{1\over T})$.

For the coupling constant $g_s$, we have:
\begin{eqnarray}\label{final-b}\nonumber
& g_s f_{-,{1\over 2} } f_{+, {3\over 2} } ={1\over 6}m_\rho^2
 e^{ { {\bar\Lambda}_{-,{1\over 2} } +{\bar \Lambda}_{+,{3\over 2} } \over T }}
 u_0\{ f^T_\rho \varphi^T_{\rho} (u_0)T(1-e^{-{\omega_c\over T}}) 
-{1\over 2} f_\rho m_\rho g_\rho^a(u_0) \} \\
&
 -{1\over 24} 
 e^{ { \Lambda_{-,{1\over 2} } +\Lambda_{+,{3\over 2} } \over T }}
\frac{d^2}{du^2}\left( 
f^T_\rho u\varphi^T_{\rho} (u) T^3f_2({\omega_c\over T})
-{1\over 2} f_\rho m_\rho ug_\rho^a(u)T^2f_1({\omega_c\over T}) 
\right)|_{u=u_0}
\;,
\end{eqnarray}
where $f_n(x)=1-e^{-x}\sum\limits_{k=0}^{n}{x^k\over k!}$ is the factor used 
to subtract the continuum. The derivative in (\ref{final-b}) arises from 
the factor $(q\cdot v)^2$ in (\ref{r7}). We have used integration by parts 
to absorb the factor $(q\cdot v)^2$. In this way we 
arrive at the simple form after double Borel transformation. 

%%%%%%%%%%%%%%%%%%%%%%%%%%%%%%%%%%%%%%%%%%%%%%%%%%%%%%%%%%%%%%%%%%%%%%%%%%%%%%%%
\section{Determination of the parameters}
\label{sec3} 
In order to obtain the coupling constants 
from (\ref{final-a}) and (\ref{final-b}) we need to use
the mass parameters $\bar\Lambda$'s and the coupling constants $f$'s of the
corresponding interpolating currents as input. $\bar\Lambda_{-,1/2}$ and $%
f_{-,1/2}$ can be obtained from the results in \cite{neubert} as $%
\bar\Lambda_{-,1/2}=0.5$ GeV and $f_{-,1/2}\simeq 0.25$ GeV$^{3/2}$ at the
leading order of $\alpha_s$. Notice that 
the coupling constant $f_{-,1/2}$ defined in
the present work is a factor $1/\sqrt 2$ smaller than that defined in \cite
{neubert}. $\bar\Lambda_{+,3/2}$ is given in \cite{huang}. $f_{+,3/2}$
can be determined from the formulas (34) of reference \cite{huang} 
derived from sum rules for two point correlators. The results are 
\begin{eqnarray}
\label{fvalue}
&&\bar\Lambda_{+,3/2}=0.82 ~~\mbox{GeV}\hspace{1.2cm}f_{+,3/2}=0.19\pm 0.03 ~~\mbox{GeV}^{5/2}\;,
\end{eqnarray}

We choose to work at the symmetric point $T_1=T_2=2T$, i.e., 
$u_0 ={1\over 2}$ as traditionally done in literature \cite{bely95}. 
The mass difference between $(1^+, 2^+)$ and $(0^-, 1^-)$
doublets is about $0.5$GeV in the leading order of HQET, which is 
much smaller than the large value of $T_1, T_2$ used below, 
$T_1, T_2\sim 3.6$GeV. So such a choice is reasonable. 
The values of the various functions appearing 
in (\ref{final-a}), at $u_0={1\over2}$, are: 
$\varphi^T_\rho(u_0)=1.552$,
$g^a_\rho(u_0)=1.5$, 
${\varphi^{\prime\prime}}^T_\rho(u_0)=-11.55$ and 
${g^{\prime\prime}}^a_\rho(u_0)=-12$. 

\section{Numerical results and discussion}
\label{sec4}
We now turn to the numerical evaluation of the sum rules for the coupling
constants. The upper limit of $T$ is constrained by the requirement that
the continuum contribution is less than $40\%$. 
This corresponds to $T<2.5$GeV. 
The lower limit of $T$ is at the point of T where stability 
develops. This leads to $T>1.0$ GeV for the sum rules (\ref{final-a}).

Stability develops for the sum rule (\ref{final-a})-(\ref{final-b})
in the region $1.0$ GeV $<$$T$$<$$2.5$ GeV.
With the values of pion wave functions 
at $u_0 ={1\over 2}$ we obtain the sum rules for 
$g_d f_{-,{1\over 2} } f_{+, {3\over 2} }$ and 
$g_s f_{-,{1\over 2} } f_{+, {3\over 2} }$
as functions of $T$ and $\omega_c$.
The results are plotted as curves in Fig. 1 and 2
with $\omega_c=3.2, 3.0, 2.8$ GeV. 

Numerically we have:
\begin{eqnarray}
\label{res-1}
 &&g_df_{-,{1\over 2} } f_{+, {3\over 2} }=(0.17\pm 0.02)
 \mbox{GeV}^2\;,
\end{eqnarray}
\begin{eqnarray}
\label{res-2}
 &&g_sf_{-,{1\over 2} } f_{+, {3\over 2} }=(0.095\pm 0.02)
  \mbox{GeV}^4 \; ,
\end{eqnarray}
where the errors refers to the variations with $T$ 
and $\omega_c$ in this region.
The central value corresponds to $T=1.8$GeV and $\omega_c =3.0$GeV.

With the values of f's in (\ref{fvalue}), we get:
\begin{eqnarray}
\label{re-1}
 &&g_d=(3.8\pm 0.4\pm 0.8) \mbox{GeV}^{-2}\;,
\end{eqnarray}
\begin{eqnarray}
\label{re-2}
 &&g_s=2.1\pm 0.4\pm 0.4 \; ,
\end{eqnarray}
where the second error takes into account the uncertainty in f's.
The inherent uncertainties due to the method of QCD sum rules and the 
choice of the rho meson wave functions are not included here. 

The decay width formulas in the leading order of HQET are 
\begin{eqnarray}
\label{widths}
&&\Gamma(B_1\to B\rho)={1\over 144\pi} g_d^2|\vec q|^5
+{1\over 8\pi} g_s^2|\vec q| \;,\nonumber\\
&&\Gamma(B_1\to B^*\rho)={1\over 16\pi} \{
\left( {14\over 9}+{1\over 27}{|\vec q|^2\over m_\rho^2} \right)
g_d^2|\vec q|^5+{2\over 9}{|\vec q|^5\over m_\rho^2}g_d g_s +
\left( 1+{1\over 3}{|\vec q|^2\over m_\rho^2} \right)
g_s^2|\vec q| \} \;,\nonumber\\
&&\Gamma(B_2^*\to B\rho)={3\over 80\pi} g_d^2|\vec q|^5\;,\nonumber\\
&&\Gamma(B_2^*\to B^*\rho)={1\over 16\pi} \{
\left( {4\over 3}+{1\over 9}{|\vec q|^2\over m_\rho^2} \right)
g_d^2|\vec q|^5+{2\over 3}{|\vec q|^5\over m_\rho^2}g_d g_s +
\left( 3+{|\vec q|^2\over m_\rho^2} \right)
g_s^2|\vec q| \} \;,
\end{eqnarray}
where $|\vec q|=\sqrt{(m_1^2-(m_2+s)^2)(m_1^2-(m_2-s)^2)}/2m_1$, 
$m_1$, $m_2$ is the parent and decay heavy meson mass, $s$ is the 
$\rho$ meson mass and $m_\rho =770$MeV is the $\rho$ meson central mass.

In (\ref{widths}) the sum over charged and neutral rho meson final states has
been included. In order to estimate the two-pion transition widths
of the $(1^+,2^+)$ doublet, we assume these transitions are dominated by the 
low-mass tail of the $\rho$ resonance. Note $2m_\pi \le s\le m_1-m_2$.

We apply the leading order formulas obtained above to the
excited states of charmed mesons assuming the HQET holds well for the charm system. 
The value $\vec q$ for
these processes in (\ref{widths}) are calculated from the experimental
mass values of the relevent particles \cite{review}. We have to smear 
the expression (\ref{widths}) with a Breit-Wigner form to take into 
account of the $150$-MeV width of the rho meson, 
$f(s^2)={1\over\pi}{m_\rho\Gamma_\rho\over (s^2-m_\rho^2)^2+m_\rho^2\Gamma_\rho^2}$.
Our final results are:
\begin{eqnarray}
\label{numerical}
&&\Gamma(D_1\to D\rho)=2.32 \mbox{MeV}\;,\nonumber \\
&&\Gamma(D_1\to D^*\rho)=0.08\mbox{MeV}\;,\nonumber\\
&&\Gamma(D_2^*\to D\rho)=0.05\mbox{MeV}\;,\nonumber\\
&&\Gamma(D_2^*\to D^*\rho)=0.51\mbox{MeV}
\end{eqnarray}

The masses of $B_1, B_2^\ast$ are not known experimentally. 
If we use $m_{B_1}=5.755$GeV and $m_{B^\ast_2}=5.767$GeV from the 
quark model \cite{eichten93}, the decay widths are:
\begin{eqnarray}
\label{num-bottom}
&&\Gamma(B_1\to B\rho)=1.11 \mbox{MeV}\;,\nonumber \\
&&\Gamma(B_1\to B^*\rho)=0.34\mbox{MeV}\;,\nonumber\\
&&\Gamma(B_2^*\to B\rho)=0.01\mbox{MeV}\;,\nonumber\\
&&\Gamma(B_2^*\to B^*\rho)=1.16\mbox{MeV}
\end{eqnarray}

In reference \cite{eichten93}, the above decay widths were estimated 
based on the potential model and the assumption that the strange quark 
satisfies the heavy quark symmetry:
\begin{eqnarray}
\label{prl1}
&&\Gamma(D_1\to D\rho)=6\mbox{MeV}\;,\nonumber \\
&&\Gamma(D_1\to D^*\rho)= < 1\mbox{MeV}\;,\nonumber\\
&&\Gamma(D_2^*\to D\rho)=< 1\mbox{MeV}\;,\nonumber\\
&&\Gamma(D_2^*\to D^*\rho)=3 \mbox{MeV}
\end{eqnarray}
\begin{eqnarray}
\label{prl2}
&&\Gamma(B_1\to B\rho)=3\mbox{MeV}\;,\nonumber \\
&&\Gamma(B_1\to B^*\rho)= 1\mbox{MeV}\;,\nonumber\\
&&\Gamma(B_2^*\to B\rho)=< 1\mbox{MeV}\;,\nonumber\\
&&\Gamma(B_2^*\to B^*\rho)=3 \mbox{MeV}
\end{eqnarray}

In this work we have calculated the two-pion transitions of the $(1^+,2^+)$
doublets through the low-mass tail of the rho resonance 
with the method of the light cone QCD sum rules in the 
leading order of $1/m_Q$. The derived two-pion 
decay widths are smaller than the estimated values in \cite{eichten93}.
Although we may conclude that the total width of 
$D^*_2$ and $D_1$ is dominated by one pion decay, 
the two-pion decay channels are not negligible. 
Experimental data is not yet available for $B_1$ and $B_2^\ast$ mesons. 
From the above discussions we know that they should also 
have significant two-pion decay widths.

%%%%%%%%%%%%%%%%%%%%%%%%%%%%%%%%%%%%%%%%%%%%%%%%%%%%%%%%%%%%%%%%%%%%%%%%%%%%%%%%

\vspace{0.8cm} {\it Acknowledgements:\/} S.Z. was supported by
the National Postdoctoral Science Foundation of China and Y.D. was supported by 
the National Natural Science Foundation of China.

\vspace{1cm}

{\bf Figure Captions}
\vspace{2ex}
\begin{center}
\begin{minipage}{130mm}
{\sf Fig. 1.} \small{ The sum rule for 
$f_{-,{1\over 2}} f_{+,{3\over 2}}g_d$ 
with $\omega_c=3.2, 3.0, 2.8$ GeV. }
\end{minipage}
\end{center}
\begin{center}
\begin{minipage}{130mm}
{\sf Fig. 2.} \small{The sum rule for 
$f_{-,{1\over 2}} f_{+,{1\over 2}}g_s$ 
with $\omega_c=3.2, 3.0, 2.8$ GeV. }
\end{minipage}
\end{center}

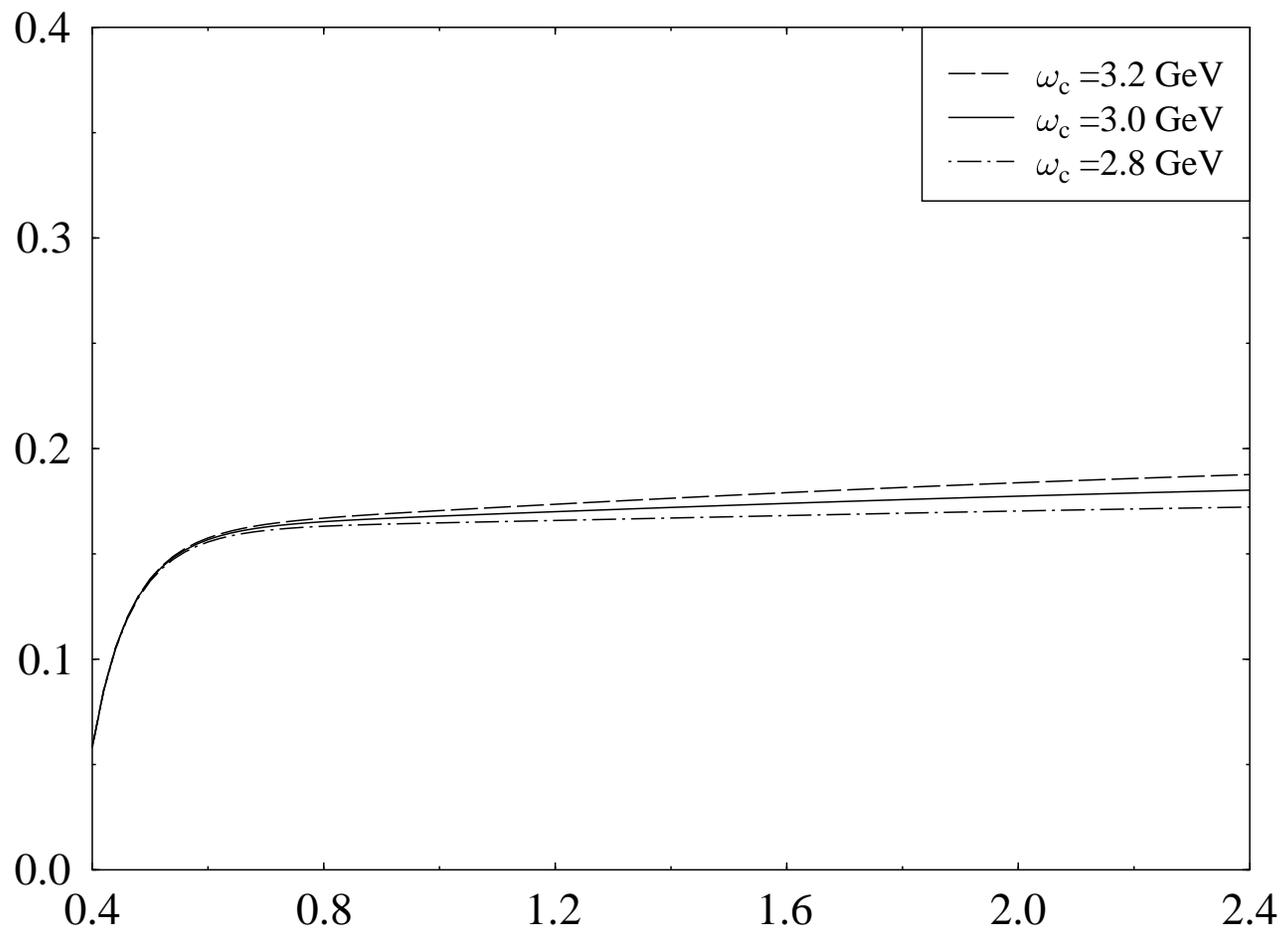
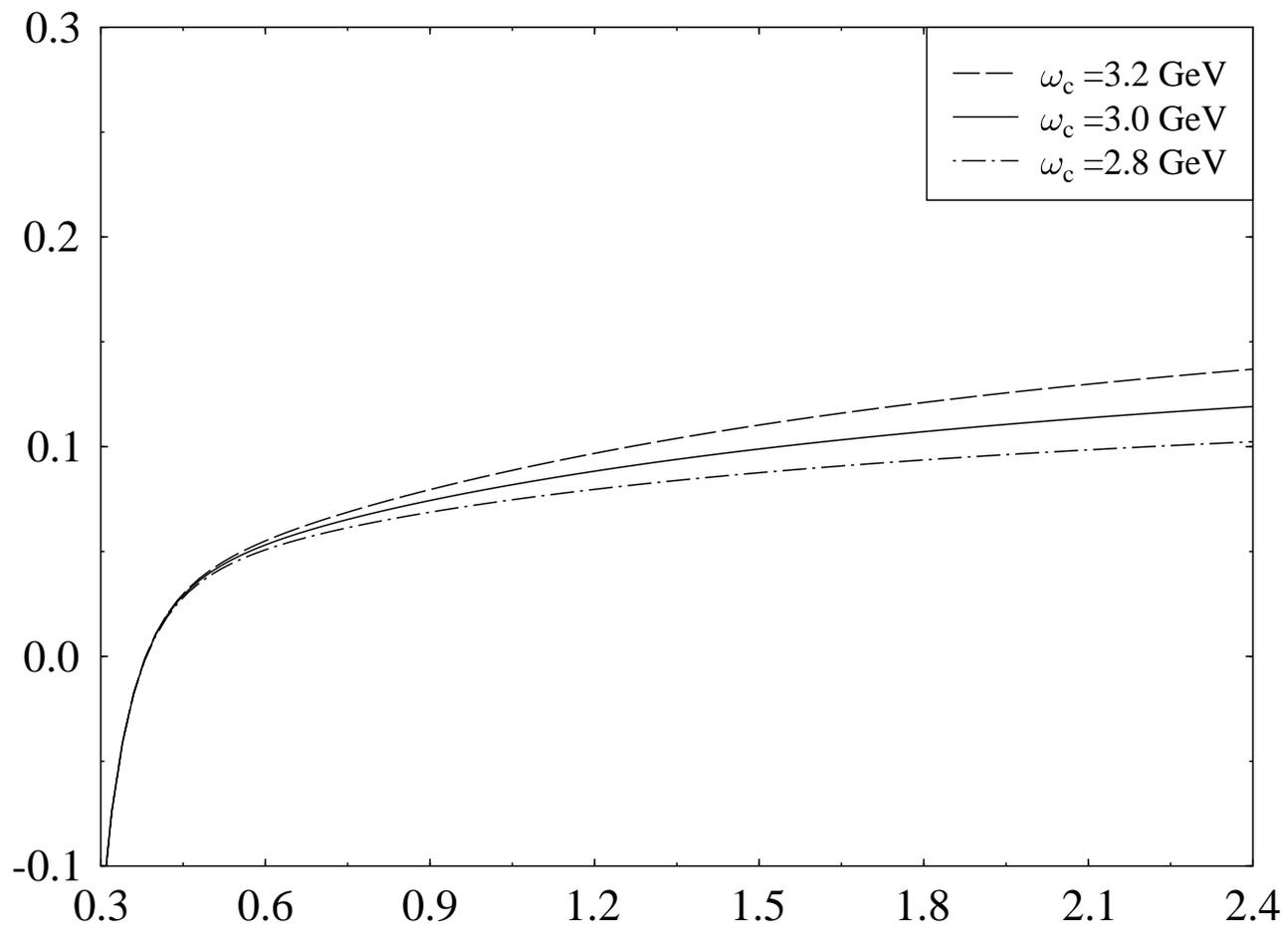
\end{document}